\documentclass[12pt,a4paper]{conference}

\usepackage{fancyhdr}
\usepackage{graphicx,amsmath,amssymb,cite}
\usepackage{multind}
\makeindex{author} \makeindex{subject}

\newcommand{\beq}{\begin{equation}}
\newcommand{\eeq}[1]{\label{#1}\end{equation}}
\newcommand{\eeqn}{\end{equation}}

\newcommand{\beqa}{\begin{eqnarray}}
\newcommand{\eeqa}[1]{\label{#1}\end{eqnarray}}
\newcommand{\eeqan}{\end{eqnarray}}

\let\bar=\overbar

\newcommand{\Dslash}{\not{\hbox{\kern-4pt $D$}}}
\newcommand{\dslash}{\not{\hbox{\kern-2pt $\del$}}}

\newcommand{\msb}{{\bar{\ssstyle M \kern -1pt S}}}

\begin{document}
%%%%%%%%%%%%%%%%%%%%%%%%%%%%%%%%%%%%%%%%%%%%%%%%%%%%%%%%%%%%%%%%%%%%%%%

\Chapter{Multichannel chiral approach for kaonic hydrogen}
           {Multichannel chiral approach for kaonic hydrogen}
           {A.~Ciepl\'{y}, J.~Smejkal}
\vspace{-6 cm}\includegraphics[width=6 cm]{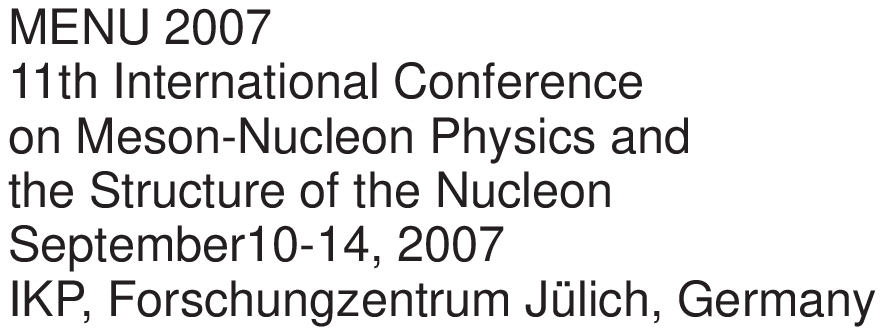}
%\bigskip\bigskip
\vspace{3.5 cm}

\addcontentsline{toc}{chapter}{{\it A. Ciepl\'{y}}} \label{authorStart}
%%%%%%%%%%%%%%%%%%%%%%%%%%%% NEW SWITCHES %%%%%%%%%%%%%%%%%%%%%%%%%%%%%%

\begin{raggedright}

{\it A.~Ciepl\'{y} 
\footnote{cieply@ujf.cas.cz}}
\index{author}{Ciepl\'{y}, A.}\\
Nuclear Physics Institute\\
250 68 \v{R}e\v{z}\\
Czech Republic
\bigskip

{\it J.~Smejkal }
\index{author}{Smejkal, J.}\\
Czech Technical University \\
Institute of Experimental and Applied Physics \\
Horsk\'{a} 3a/22 \\
128 00 Praha 2 \\
Czech Republic
\bigskip

\end{raggedright}

\begin{center}
\textbf{Abstract}
\end{center}
We present an exact solution to the $K^{-}$-proton bound state problem 
formulated in the momentum space. The 1s level characteristics of 
the kaonic hydrogen are described together with the available 
low energy $\bar{K}N$ data.

\section{Introduction}

We developed a precise method of computing the meson-nuclear bound
states in momentum space. The method was already applied to
pionic atoms and its multichannel version was used to calculate the 1s level
characteristics of pionic hydrogen \cite{96CiM}. Here we just 
remark that our approach is based on the construction of the Jost matrix 
and involves the solution of the Lippman-Schwinger equation for the transition 
amplitudes between the various channels. In this work we aim at simultaneous 
description of both the 1s level kaonic bound state and the available 
experimental data for the $K^{-}p$ initiated processes. 

In view of the vastly improved experimental results on the 1s level 
of kaonic hydrogen \cite{05DEAR} the exact solution of the bound state 
problem should be preferred over the traditional Deser-Trueman formula 
relating the threshold scattering amplitude to the hadronic energy level 
characteristics of exotic atoms. Recently, the relation 
for kaonic hydrogen was modified to include the isospin effects 
and electromagnetic corrections \cite{04MRR}.

\section{Meson-baryon potentials}

Unlike the pion-nucleon interaction the $\bar{K}N$ dynamics is strongly 
influenced by the existence of the $\Lambda(1405)$ resonance, just below 
the $K^{-}p$ threshold. This means that the standard chiral perturbation 
theory is not applicable in this region. Fortunately, one can use 
non-perturbative coupled channel techniques to deal with the problem and 
generate the $\Lambda(1405)$ resonance dynamically. Such approach has proven 
quite useful and several authors have already applied it to various low 
energy meson-baryon processes \cite{95KSW, 05BNW, 06Oll}. 

Here we follow the approach of Ref.~\cite{95KSW} and take the strong 
interaction part of the potential matrix in a separable form 
\beq
V_{ij}(k,k')=\sqrt{\frac{1}{2E_i}\frac{M_i}{\omega_i}} \: 
g_{i}(k)\:\frac{C_{ij}}{f^2} \:g_{j}(k')
\:\sqrt{\frac{1}{2E_j}\frac{M_j}{\omega_j}},
\quad g_{j}(k)=\frac{1}{1+(k/ \alpha_{j})^2}
\eeq{poten}
in which the parameter $f$ stands for the pseudoscalar meson decay constant 
in the chiral limit. The coupling matrix $C_{ij}$ is determined by 
chiral SU(3) symmetry and it includes terms up to second order in the meson 
c.m. kinetic energies. The off shell form factor $g_{j}(k)$ introduces 
the inverse range radius $\alpha_{j}$ that characterizes the radius of 
interaction in the channel $j$. In the Born approximation the potentials 
$V_{ij}(k,k')$ give the same (up to $\mathcal{O}(q^2)$) s-wave scattering 
lengths as are those derived from the underlying chiral lagrangian. More 
details on the construction of the effective (chirally motivated) potentials 
and the specification of the kinematical factors $\sqrt{M_j/(2E_j \omega_j)}$ 
can be found in Refs.~\cite{95KSW, 07CSm}. While the authors of Ref.~\cite{95KSW} 
restricted themselves only to the six channels that are open at the $\bar{K}N$ 
threshold we have employed all ten coupled meson-baryon channels in our model: 
$K^-p$, $\bar{K}^{0}n$, $\pi^{0}\Lambda$, $\pi^{+}\Sigma^{-}$, 
$\pi^{0}\Sigma^{0}$, $\pi^{-}\Sigma^{+}$, $\eta \Lambda$, 
$\eta \Sigma^0$, $K^+\Xi^-$, and $K^0 \Xi^0$. The potential of Eq.~(\ref{poten}) 
is used not only when solving the bound state problem but we also implement 
it in the standard Lippman-Schwinger equation and compute the low energy 
$\bar{K}N$ cross sections and branching ratios from the resulting 
transition amplitudes.

\section{$\bar{K}N$ data fits}

The parameters of the chiral lagrangian that enter the coefficients 
$C_{ij}$ and the inverse range radii $\alpha_{i}$ determining the off-shell 
behavior of the potentials are to be fitted to the experimental data. Before 
performing the fits we reduce the number of the fitted parameters 
in the following way. First, the axial couplings $D$ and $F$ (concerning 
the specification of the various chiral couplings we refer the reader 
to Refs.~\cite{95KSW} and \cite{07CSm}) have already been established 
in the analysis of semileptonic hyperon decays, $D = 0.80$, \mbox{$F = 0.46$} 
($g_{A} = F + D = 1.26$). Then, we fix the couplings $b_0$ and $b_F$ 
to satisfy the approximate Gell-Mann formulas for the baryon mass splittings, 
$b_0 = 0.064$~GeV$^{-1}$ and $b_F = -0.209$~GeV$^{-1}$. Similarly, we determine 
the coupling $b_0$ and the baryon chiral mass $M_0$ from the relations 
for the pion-nucleon sigma term $\sigma_{\pi N}$ and for the proton mass 
(see e.g.~\cite{06Oll}). Finally, we reduce the number of the inverse 
ranges $\alpha_{i}$ to only five: $\alpha_{KN}$, $\alpha_{\pi \Lambda}$, 
$\alpha_{\pi \Sigma}$, $\alpha_{\eta \Lambda /\Sigma}$, 
$\alpha_{K \Xi}$. This leaves us with 11 free parameters: the five 
inverse ranges, the pseudoscalar meson decay constant $f$, and five more 
couplings from the second order chiral lagrangian denoted by 
$d_D$, $d_F$, $d_0$, $d_1$, and $d_2$.

The fits to low energy $\bar{K}N$ data standardly include the three 
precisely measured threshold branching ratios $\gamma$, $R_c$ and $R_n$  
(specified e.g. in Ref.~\cite{95KSW}) and the $K^- p$-initiated total 
cross sections. For the later ones we consider only the experimental data 
taken at the kaon laboratory momenta \mbox{$p_{LAB} = 110$~MeV}
(for the $K^- p$, $\bar{K^0}n$, $\pi^{+} \Sigma^{-}$,
$\pi^{-} \Sigma^{+}$ final states) and at $p_{LAB} = 200$~MeV (for 
the same four channels plus $\pi^0 \Lambda$ and $\pi^{0} \Sigma^{0}$). 
Our results show that the inclusion of more data taken at other kaon momenta 
is not necessary since the fit at just $1-2$ points fixes the cross section 
magnitude and the energy dependence is reproduced nicely by the model. 
With the inclusion of the DEAR results on the strong interaction shift 
$\Delta E_N$ and the width $\Gamma$ of the 1s level in kaonic hydrogen 
we end up with a total of 15 data points in our fits.

\vspace*{2mm}
{\bf Table 1:} The fitted $\bar{K}N$ threshold data \\[2mm]
\begin{tabular}{ccccccc}
$\sigma_{\pi N}$ [MeV] & $\chi^{2}/N$ & $\Delta E_{N}$ [eV] & $\Gamma$ [eV]& $\gamma$ & $R_c$ & $R_n$ \\ \hline
 20                & 1.33  & 232     & 725      & 2.366   & 0.657 & 0.191 \\
 30                & 1.36  & 272     & 683      & 2.367   & 0.658 & 0.190 \\
 40                & 1.38  & 257     & 713      & 2.370   & 0.658 & 0.190 \\
 50                & 1.40  & 266     & 708      & 2.370   & 0.658 & 0.190 \\ \hline
 exp               &   -   & 193(43) & 249(150) & 2.36(4) & 0.664(11) & 0.189(15)
\end{tabular}

\vspace*{2mm}
Our results are summarized in Table 1, where the results of our 
$\chi^{2}$ fits are compared with the relevant experimental data. 
Since the value of the pion-nucleon $\sigma$-term is not well determined we 
enforced four different options, which cover the interval of the values 
considered by various authors. The resulting $\chi^{2}$ per data point 
indicate satisfactory fits. It is worth noting that their quality and the 
computed values do not depend much on the exact value of the $\sigma_{\pi N}$. 
The low energy cross sections included in the fits are not shown here but we 
stress that their description is good \cite{07CSm}. The strong interaction 
energy shift of the 1s level in kaonic hydrogen is reproduced well but we 
were not able to get a satisfactory fit of the 1s level energy width as our 
results are significantly larger than the experimental value. However, when 
considering the interval of three standard deviations and also the older KEK 
results (that give less precise but larger width) one cannot conclude that 
kaonic hydrogen measurements contradict the other low energy 
$\bar{K}N$ data. 

In Table 2 we compare our results (for $\sigma_{\pi N}=20,\:30$ 
and $50$ MeV) for the 1s level characteristics in kaonic hydrogen 
with the approximate values determined from the $K^{-}p$ 
scattering lengths $a_{K^{-}p}$ that were obtained from the multiple 
channel calculation that uses the same parametrization of the strong 
interaction potential (\ref{poten}). The 1s level complex energies are 
shown for: the standard Deser-Trueman formula (DT), the modified 
Deser-Trueman formula (MDT) \cite{04MRR} and our ``exact'' solution 
of the bound state problem. We have checked that if only the point-like 
Coulomb potential is considered in the $K^{-}p$ channel our method 
reproduces the well known Bohr energy of the 1s level with a precision 
better than $0.1$~eV. This means that the discrepancy between the ``MDT'' and the 
``exact'' values can be attributed to higher order corrections not 
considered in the derivation of the MDT. In view of the current level 
of the experimental precision the use of the MDT formula is sufficient. 
Though, the situation may change after the coming SIDDHARTA experiment.

\vspace*{2mm}
{\bf Table 2:} Precision of the Deser-Trueman formula\\[2mm]
\begin{tabular}{cccc}
$a_{K^{-}p}$ [fm]   & $-0.50 + {\rm i}\:1.01$ & $-0.59 + {\rm i}\:0.99$ & $-0.60 + {\rm i}\: 1.01$ \\ \hline  
DT:  $\Delta E_{N} - ({\rm i}/2)\Gamma$ [eV]   & $207 - ({\rm i}/2)832$ & $256 - ({\rm i}/2)806$ & $247 - ({\rm i}/2)830$ \\
MDT: $\Delta E_{N} - ({\rm i}/2)\Gamma$ [eV]   & $251 - ({\rm i}/2)714$ & $290 - ({\rm i}/2)664$ & $285 - ({\rm i}/2)689$ \\
exact: $\Delta E_{N} - ({\rm i}/2)\Gamma$ [eV] & $232 - ({\rm i}/2)725$ & $262 - ({\rm i}/2)698$ & $266 - ({\rm i}/2)708$
\end {tabular}

\vspace*{4mm}
{\bf Acknowledgement:} A.~C. acknowledges a financial support from the GA~AVCR 
grant A100480617.

%\begin{thebibliography}{000} %for 3 digits
%\begin{thebibliography}{00}  %for 2 digits

%%%%%%%%%%%%%%%   Author and Subject Index
\printindex{author}{Author Index}
\blankpage

\printindex{subject}{Subject Index}
\blankpage

\begin{thebibliography}{0}    %for 1 digit

%%journal paper
\bibitem{96CiM} A. Ciepl\'{y} and R. Mach, {\it Nucl. Phys.} {\bf A609}, 377 (1996)

\bibitem{05DEAR}  G. Beer {\it et al}.~[DEAR Collab.],
{\it Phys. Rev. Lett.} {\bf 94}, 212302 (2005).

\bibitem{04MRR} U.-G.~Meissner, U.~Raha, and A.~Rusetsky, {\it Eur. Phys. J.} 
  {\bf C35}, 349 (2004)

\bibitem{95KSW} N.~Kaiser, P.B.~Siegel, and W.~Weise, {\it Nucl. Phys.} 
  {\bf A594}, 325 (1995)

\bibitem{05BNW} B.~Borasoy, R.~Nissler, and W.~Weise, {\it Eur. Phys. J.} 
  {\bf A25}, 79 (2005)
  
\bibitem{06Oll} J.A.~Oller, {\it Eur. Phys. J.} {\bf A28}, 63 (2006)

\bibitem{07CSm} A.~Ciepl\'{y} and J.~Smejkal - in preparation

\end{thebibliography}
\end{document}